\documentclass[a4paper,11pt]{article}
\pdfoutput=1 

\usepackage{jcappub} 

\usepackage[T1]{fontenc} 
\usepackage{graphicx}
\usepackage{dcolumn}
\usepackage{bm}
\usepackage{amsmath, amssymb}
\usepackage{upgreek}
\usepackage{verbatim}
\usepackage{hyperref}
\newcommand{\lcdm}{$\Lambda$CDM}
\newcommand{\om}{\Omega_{m0}}
\newcommand{\obh}{\Omega_{b0} h^2}
\newcommand{\och}{\Omega_{c0} h^2}

\title{Constraints on power law cosmology from cosmic chronometer, standard ruler, and standard candle data}

\author[a]{Joseph Ryan}

\affiliation[a]{Department of Physics, Kansas State University, 116 Cardwell Hall, 1228 N. Martin Luther King Jr. Drive, Manhattan, KS 66506, USA}

\emailAdd{jwryan@ksu.edu}

\abstract{In this paper I investigate how well simple power law expansion fits observational data in comparison to the standard \lcdm\ model. I analyze a data set consisting of cosmic chronometer, standard ruler, and standard candle measurements, finding that the \lcdm\ model provides a better fit to most combinations of these data than the power law ansatz.

\vspace{2mm}

\noindent keywords: dark energy experiments, dark energy theory}

\begin{document}
\maketitle
\flushbottom

\section{Introduction}

There is now broad consensus that the \lcdm\ model
adequately describes the dynamics of the Universe, on large scales, throughout most of its history \cite{Planck_overview}. Some groups \cite{Cao_Ryan_Khadka_Ratra, Cao_Ryan_Ratra, Ryan_2, Ryan_1, Ooba_Ratra_Sugiyama_2017_NFLCDM, Park_Ratra_2018_FLCDM, Ooba_Ratra_Sugiyama_2017_NFpCDM, Ooba_Ratra_Sugiyama_2017_NFXCDM, Park_Ratra_2018_FXCDM_NFXCDM, Park_Ratra_2018_FpCDM_NFpCDM, Ooba_Ratra_Sugiyama_2018_FpCDM, zhang_et_al_2017a, wang_pogosian_zhao_zucca_2018, zhang_lee_geng_2018, Handley_2019, Di_Valentino_et_al_2020, Luo_et_al_2020} have found evidence of departures from the \lcdm\ model (such as dynamical dark energy or large-scale spatial curvature), but these findings have generally not risen to the level of significance necessary to unseat \lcdm\ from its position as the standard model of cosmology.\footnote{The evidence of large-scale spatial curvature, in particular, has recently been challenged by an analysis of Hubble parameter data \cite{Vagnozzi_et_al_2021}.} Instead of introducing alternative cosmological models, a more direct way to look for cracks in the standard model is to constrain the form of the scale factor, $a(t)$, independently of a specific model, using observational data. 
The \lcdm\ model makes a definite, testable prediction of the scale factor's evolution with cosmic time $t$, and while this evolution appears to be in good agreement with the available data, one could ask whether a simpler expansion history may describe these data equally well or better. For example, one could propose that the scale factor take a power law form $a(t) \propto t^{\beta}$, where $\beta$ is a constant exponent. This ansatz has many virtues, one of which is its simplicity. Power law expansion only depends on the single parameter $\beta$, and the functional form $t^{\beta}$ is easy to integrate analytically when it appears in the integral $\int \frac{dt}{a(t)}$ (as in the computation of the co-moving distance scale). Additionally, power law expansion with $\beta \geq 1$ has neither a horizon problem nor a flatness problem, and produces a universe whose age is compatible with the ages of the oldest known objects in the Universe (these being globular clusters and high-redshift galaxies \cite{Dev_Jain_Lohiya_2008, Sethi_Dev_Jain_2005}). 
Power law expansion is also a predicted feature of some alternative gravity theories that are designed to solve the cosmological constant problem \cite{Dev_Jain_Lohiya_2008, Sethi_Dev_Jain_2005}. Many investigators have found that $\beta \approx 1$ is favored by various independent low-redshift probes, such as cosmic chronometers ($H(z)$) \cite{Dev_Jain_Lohiya_2008}, gravitational lensing statistics \cite{Dev_Safonova_Deepak_Lohiya_2002}, Type Ia supernovae (SNe Ia) \cite{Dev_Sethi_Lohiya_2001, Kumar_2012, Rani_et_al_2015, Sethi_Dev_Jain_2005}, baryon acoustic oscillations (BAO) \cite{Shafer_2016, Tutusaus_et_al_2016}, quasar angular sizes (QSO) \cite{Jain_Dev_Alcaniz_2003}, galaxy cluster gas mass fractions \cite{Zhu_Alcaniz_Liu_2008}, and the combination of $H(z)$ + BAO + SNe Ia + gamma-ray burst distance moduli (GRB) \cite{Haridasu_AAP_2017}. Other data sets, however, favor $\beta \approx$ 1.2-1.6 \cite{Dev_Jain_Lohiya_2008, Dolgov_Halenka_Tkachev_2014, Kumar_2012, Shafer_2016, Tutusaus_et_al_2016}; see Table \ref{tab:beta_fits}.

Some studies have also found that power law expansion with $\beta = 1$ can produce the right amount of primordial helium to match current observations \cite{Lohiya_Batra_Mahajan_Mukherjee_1999, Sethi_Batra_Lohiya_1999}, and so may be able to account for the synthesis of other light elements. These conclusions are challenged by the results of other studies, which find that $\beta \approx$ 0.55-0.58 is required to produce the right abundances \cite{Kaplinghat_Steigman_Walker_2000, Kaplinghat_Steigman_Tkachev_1999, Kumar_2012}. If these latter studies are correct, then the values of $\beta$ favored by primordial nucleosynthesis are clearly disjoint with those favored by low redshift measurements, and it is difficult to see how they can be reconciled without introducing extra complexity to the power law ansatz (such as the addition of a mechanism that forces $\beta$ to change its value between the two eras; see e.g. \cite{Gumjudpai, Gumjudpai_Thepsuriya_2012, Kaeonikhom_Gumjudpai_Saridakis_2011, Rangdee_Gumjudpai_2014, Wei_2004}).\footnote{For recent efforts to provide an account of primordial nucleosynthesis within a power law ansatz, see \cite{Singh_Lohiya_2015_arXiv, Singh_Lohiya_2015_JCAP}.} Given that power law expansion (whatever underlying cosmological model may be required to produce it) is intended to offer a simpler alternative to the expansion history predicted by the \lcdm\ model, such additional complexity seems unjustified, and the power law ansatz with constant $\beta$ appears to be ruled out on these grounds.

\begin{table*}
    \centering
    \caption{Fits to power law exponent from other low redshift measurements.}
    \begin{tabular}{ccc}
    \hline
    \hline
         Reference & $\beta$ & Data type(s) used \\
         \hline
         \cite{Dev_Jain_Lohiya_2008} & $1.07^{+0.11}_{-0.08}$ & $H(z)$\\
         & $1.42^{+0.08}_{-0.07}$ & SN Ia\\
         \cite{Dev_Safonova_Deepak_Lohiya_2002} & $1.09\pm0.3$ & Gravitational lensing statistics\\
         & $1.13^{+0.4}_{-0.3}$ & \\
         \cite{Dev_Sethi_Lohiya_2001} & $1.004\pm0.043$ & SN Ia\\
         \cite{Dolgov_Halenka_Tkachev_2014} & $1.52 \pm 0.15$ & SN Ia\\
         & $1.55 \pm 0.13$ & \\
         & $1.3$ & BAO\\
         \cite{Jain_Dev_Alcaniz_2003} & $1.0\pm0.3$ & QSO\\
         \cite{Kumar_2012} & $1.22_{-0.16}^{+0.21}$ & $H(z)$\\
         & $1.61_{-0.12}^{+0.14}$ & SN Ia\\
         \cite{Rani_et_al_2015} & $1.05_{-0.066}^{+0.071}$ & $H(z)$\\
         & $1.44^{+0.26}_{-0.18}$ & SN Ia\\
         \cite{Sethi_Dev_Jain_2005} & $1.04^{+0.07}_{-0.06}$ & SN Ia\\
         \cite{Shafer_2016} & $0.93$ & BAO\\
         & $1.44$-$1.56$ & SN Ia\\
         \cite{Tutusaus_et_al_2016} & $0.908\pm0.019$ & BAO\\
         & $1.55\pm0.13$ & SN Ia\\
         \cite{Zhu_Alcaniz_Liu_2008} & $1.14\pm0.05$ & Galaxy cluster gas mass fraction\\
         \cite{Haridasu_AAP_2017} & $1.08 \pm 0.04$ & $H(z)$ + BAO + SNe Ia + GRB\\
    \hline
    \hline
    \end{tabular}
    \label{tab:beta_fits}
\end{table*}

A defender of power law expansion who does not wish to make the expansion history more complex by introducing a time-variable $\beta$ could attempt to save it by arguing that:

1.) The findings of \cite{Kaplinghat_Steigman_Walker_2000, Kaplinghat_Steigman_Tkachev_1999, Kumar_2012} are simply incorrect, and the power law exponent has the value $\beta \approx 1$ during both the nucleosynthesis era and the present era, or

2.) The Universe only undergoes power law expansion at late times, and the power law ansatz with $\beta \approx 1$ adequately describes low redshift observations only.

The latter option is, on its face, plausible. After all, the standard \lcdm\ model holds that the Universe follows power law expansion during both the matter-dominated and radiation-dominated eras, so it might be reasonable to limit the scope of the power law ansatz by suggesting that it only applies after the era of nucleosynthesis.\footnote{In \cite{Kolb_1989}, one of the earliest papers on the subject, the author proposes that, if a hypothetical form of matter called ``K-matter'' were to dominate the energy budget at late times, this would lead to a ``coasting'' cosmic expansion with $\beta = 1$ (with $\beta$ taking on different values in earlier eras).} We must be careful not to push this argument too far, however, because any scale factor $a(t)$ can presumably be approximated by a power law over some arbitrarily short time period. What is at issue is not whether the Universe follows power law expansion during some (relatively) brief portion or portions of its history, but whether it follows power law expansion throughout all (or most) of its history. If it can be shown that the power law ansatz fits low redshift observational data as well as or better than \lcdm\ over an appreciable range of redshifts, then option (2) is validated (and option 1 may be validated as well, if one can marshal a strong argument against the findings of \cite{Kaplinghat_Steigman_Walker_2000, Kaplinghat_Steigman_Tkachev_1999, Kumar_2012}). If, on the other hand, the power law model fails to provide a good fit to the available low redshift data, then both (1) and (2) are falsified.

A few studies \cite{Rani_et_al_2015, Shafer_2016, Tutusaus_et_al_2016, Haridasu_AAP_2017} have been conducted along these lines. These studies find that, when power law expansion is tested using multiple independent data sets ($H(z)$ alone and $H(z)$ + BAO + SNe Ia + CMB in \cite{Rani_et_al_2015}, BAO + SNe Ia in \cite{Shafer_2016}, BAO + SNe Ia + CMB in \cite{Tutusaus_et_al_2016}, and $H(z)$ + BAO + SNe Ia + GRB in \cite{Haridasu_AAP_2017}), it performs poorly compared to \lcdm. Here I continue in this vein by fitting the expansion histories predicted by the power law ansatz and the \lcdm\ model to a data set consisting of cosmic chronometer, standard ruler, and standard candle data,  some of which have not yet been used to test power law expansion (see Sec. \ref{sec:Data} for a description of the data).
I use simple model comparison statistics (the same as those used in \cite{Rani_et_al_2015, Shafer_2016, Tutusaus_et_al_2016, Haridasu_AAP_2017}; see Sec. \ref{sec:Methods}) to compare the quality of the fit in both cases. I discuss my results in Sec. \ref{sec:Results} and draw my conclusions in Sec. \ref{sec:Conclusion}.

\section{Theory}
Under the power law ansatz, the scale factor $a(t)$ takes the form
\begin{equation} \label{eq:a}
    a(t) = kt^{\beta},
\end{equation}
where $k$ and $\beta$ are constants. From the definition of redshift, $\frac{a_0}{a(t)} := 1 + z$ ($a_0$ is the current value of the scale factor and $z$ is the redshift) and eq. (\ref{eq:a}), we can write
\begin{equation}
    \frac{a_0}{kt^{\beta}} = 1 + z,
\end{equation}
from which it follows that
\begin{equation} \label{eq:1/t}
    \frac{1}{t} = \left[\frac{k}{a_0}\left(1 + z\right)\right]^{1/\beta}.
\end{equation}
The definition of the Hubble parameter, $H(t) := \frac{\dot{a(t)}}{a(t)}$, with the overdot denoting the time derivative, implies $H(t) = \frac{\beta}{t}$. Therefore eq. (\ref{eq:1/t}) can be written in the form
\begin{equation}
\label{eq:Hz_PL}
    H(z) = H_0\left(1 + z\right)^{1/\beta},
\end{equation}
where I have defined the present value of the Hubble constant to be $H_0 := \beta\left(\frac{k}{a_0}\right)^{1/\beta}$. The power law ansatz therefore has two free parameters: $H_0$ and $\beta$. 

My fiducial model in this paper is the spatially flat \lcdm\ model. In this model, at late times, the Hubble parameter can be written as a function of the redshift $z$ in the form
\begin{equation}
\label{eq:Hz_LCDM}
    H(z) = H_0 \sqrt{\Omega_{m0}\left(1 + z\right)^3 + 1 - \Omega_{m0}},
\end{equation}
where $H_0$ is the Hubble constant and $\Omega_{m0}$ is the current value of the non-relativistic matter density. The \lcdm\ model also has two free parameters: $H_0$ and $\Omega_{m0}$. Because of the relatively low redshifts of the data I use (see Table \ref{tab:Data}) I neglect the contribution that radiation makes to the energy budget.

The data that I use depend on several kinds of distance measurements (see Sec. \ref{sec:Data}). These are the Hubble distance
\begin{equation}
    \label{eq:D_H}
    D_{\rm H}(z) = \frac{c}{H(z)},
\end{equation}
the transverse co-moving distance
\begin{equation}
\label{eq:D_M}
    D_{\rm M}(z) = \frac{c}{H_0}\int^z_0 \frac{dz'}{E(z')},
\end{equation}
where $E(z) := H(z)/H_0$, the angular diameter distance
\begin{equation}
    \label{eq:D_A}
    D_{\rm A}(z) = \frac{D_{\rm M}(z)}{1 + z},
\end{equation}
the volume-averaged angular diameter distance
\begin{equation}
    \label{eq:D_V}
    D_{\rm V}(z) = \left[\frac{cz}{H_0}\frac{D_{\rm M}^2(z)}{E(z)}\right]^{1/3},
\end{equation}
and the luminosity distance
\begin{equation}
    \label{eq:D_L}
    D_{\rm L}(z) = (1 + z)D_{\rm M}(z),
\end{equation}
as defined in \cite{Farooq_thesis, Hogg}. Note that $D_{\rm M}(z)$ only has the form shown in eq. (\ref{eq:D_M}) in the special case that the Universe is spatially flat on large scales, which I assume in this paper; for open and closed universes the integral on the right-hand side is more complicated.

\section{Data}
\label{sec:Data}

\begin{table*}
    \centering
    \caption{Data used in this paper.}
    \begin{tabular}{ccc}
    \hline
    \hline
         Data type & Number of data points & Redshift range\\
         \hline
         $H(z)$ & 31 & $0.070 \leq z \leq 1.965$ \\
         BAO & 11 & $0.38 \leq z \leq 2.334$\\
         QSO & 120 & $0.462 \leq z \leq 2.73$\\
         GRB & 119 & $0.48 \leq z \leq 8.2$\\
         HIIG & 153 & $0.0088 \leq z \leq 2.42935$\\
         \hline
         \hline
    \end{tabular}
    \label{tab:Data}
\end{table*}

In Table \ref{tab:Data} I list the types of measurements I use, the number of measurements of each type, and the redshift ranges within which the measurements lie. The cosmic chronometer data consist of measurements of the Hubble parameter as a function of the redshift $z$ ($H(z)$), taken from \cite{Simon_Verde_Jimenez_2005, Stern_et_al_2010, Moresco_et_al_2012, Zhang_et_al_2014, Moresco_2015, Moresco_et_al_2016, Ratsimbazafy_et_al_2017}, and listed in \cite{Ryan_1}; see that paper for more details. To fit the power law and \lcdm\ expansion histories to the cosmic chronometer data, I compute $H(z)$ theoretically using eqs. (\ref{eq:Hz_PL}) and (\ref{eq:Hz_LCDM}).

I use two sets of standard ruler measurements in this paper. The first set consists of measurements of the quantities $H(z)$, $D_{\rm H}(z)$, $D_{\rm M}(z)$, $D_{\rm A}(z)$, and $D_{\rm V}(z)$, defined in eqs. (\ref{eq:Hz_PL}-\ref{eq:D_V}), and scaled by the value that the sound horizon $r_{\rm s}$ takes at the baryon drag epoch. These measurements are the same as those listed in Table 1 of \cite{Cao_Ryan_Ratra_2021}. See that paper, as well as \cite{Cao_Ryan_Khadka_Ratra, Cao_Ryan_Ratra, Ryan_2, Ryan_1} for more details, and for references to the original literature. To compute the size of the sound horizon, I use the approximate formula
\begin{equation}
\label{eq:r_s}
    r_{\rm s}=55.154\frac{{\rm exp}[-72.3(\Omega_{\nu 0}h^2+0.0006)^2]}{(\Omega_{\rm b 0}h^2)^{0.12807}(\Omega_{m0}h^2 - \Omega_{\nu 0}h^2)^{0.25351}} \hspace{1mm}{\rm Mpc},
\end{equation}
where $\Omega_{m0}$, $\Omega_{b0}$, and $\Omega_{\nu 0}$ are the dimensionless energy density parameters of non-relativistic matter, of baryons, and of neutrinos, respectively, and $h := H_0/100$ km s$^{-1}$ Mpc$^{-1}$ \cite{Aubourg_et_al_2015}. Following \cite{Carter_2018}, I set $\Omega_{\nu 0} = 0.0014$, which leaves two additional free parameters ($\Omega_{m0}$ and $\Omega_{b0}h^2$) when power law expansion is fitted to data combinations containing BAO data, and one additional free parameter ($\Omega_{b0}h^2$) when the \lcdm\ model is fitted to the same data combinations. The second set of standard ruler data consists of measurements of the angular sizes $\theta_{\rm obs}$, in milliarcseconds (mas), of intermediate-luminosity quasars (QSO). The angular size of a quasar can be computed theoretically via
\begin{equation}
    \label{eq:th_th}
    \theta_{\rm th}(z) = \frac{l_{\rm m}}{D_{\rm A}(z)},
\end{equation}
where $D_{\rm A}(z)$ is given by eq. (\ref{eq:D_A}) and $l_{\rm m} = 11.03 \pm 0.25$ pc is the characteristic linear size of the quasars in the sample. This quantity can then be compared to $\theta_{\rm obs}$ to determine the quality of the fit of the given expansion history to the QSO data. The QSO angular size measurements are listed, and $l_{\rm m}$ is determined, in \cite{Cao_et_al2017b}; see that paper and \cite{Ryan_2} for discussion and details.

I use two sets of standard candle data in this paper. The first set consists of measurements of the luminosities, fluxes, and velocity dispersions of HII starburst galaxies (HIIG), from which the distance moduli of these galaxies can be computed. The HIIG data consist of a low redshift ($0.0088 \leq z \leq 0.16417$) set of 107 measurements from \cite{Chavez_2014}, plus a high redshift ($0.636427 \leq z \leq 2.42935$) set of 46 measurements from \cite{G-M_2019}. Subsets of these data, which were generously provided to me by Ana Luisa Gonz\'{a}lez-Mor\'{a}n,\footnote{Private communications, 2019 and 2020.} have been used in several studies to constrain cosmological parameters \cite{Cao_Ryan_Ratra, Cao_Ryan_Ratra_2021, Cao_Ryan_Khadka_Ratra, Chavez_2012, Chavez_2016, G-M_2019, Terlevich_2015}. See \cite{Cao_Ryan_Ratra} for a detailed description of how the distance modulus can be computed. Briefly, if one knows the luminosity $L$, flux $f$, and velocity dispersion $\sigma$ of an HII galaxy, one can use these quantities to compute a distance modulus $\mu_{\rm obs}$. This quantity can then be compared to the theoretical distance modulus
\begin{equation}
\label{eq:mu_theo}
    \mu_{\rm th} = 5{\rm log}D_{\rm L}(z) + 25,
\end{equation}
where $D_{\rm L}(z)$ is given by eq. (\ref{eq:D_L}), to determine the quality of the fit of the given expansion history to the data. The second set of standard candle data consists of measurements of the bolometric fluence $S_{\rm bolo}$ and observed peak energy $E_{\rm p, obs}$ of 119 gamma-ray bursts from \cite{Dirirsa_2019} (GRB). Given a knowledge of the bolometric fluence of a source, one can compute the energy radiated isotropically in the source's rest frame
\begin{equation}
    \label{eq:E_iso}
    E_{\rm iso} = \frac{4\pi D_{\rm L}^2}{1 + z}S_{\rm bolo}.
\end{equation}
GRBs can be standardized through the Amati relation \cite{Amati_2009, Amati_2008}
\begin{equation}
\label{eq:Amati}
    {\rm log} E_{\rm iso} = a + b{\rm log}\left[\left(1 + z\right)E_{\rm p, obs}\right],
\end{equation}
which connects the observed peak energy of a given GRB to its isotropic radiated energy (here $a$ and $b$ are free parameters which I vary when fitting the power law and \lcdm\ expansion histories to the GRB data). The GRB likelihood function also contains a parameter which describes the extrinsic scatter of the GRBs in the sample ($\sigma_{\rm ext}$) \cite{D'Agostini_2005}. As in \cite{Cao_Ryan_Khadka_Ratra}, I vary this parameter freely when fitting the power law and \lcdm\ expansion histories to the GRB data. By comparing the value of ${\rm log} E_{\rm iso}$ as computed from eq. (\ref{eq:E_iso}) to that computed from eq. (\ref{eq:Amati}), one can determine the quality of the ansatz or model fit. For more details about the GRB analysis, see \cite{Cao_Ryan_Khadka_Ratra, Khadka_Ratra_2020}.

There is some overlap between the cosmic chronometer data I use in this paper and those that were used in \cite{Dev_Jain_Lohiya_2008, Kumar_2012, Rani_et_al_2015} to constrain the parameters of simple (constant $\beta$) power law expansion. Many of the measurements these authors used are the same as mine, although I use a larger, more up-to-date set (which is the same as the set of $H(z)$ data used to constrain the simple power law ansatz in \cite{Haridasu_AAP_2017}, though I add one point from \cite{Ratsimbazafy_et_al_2017}). I use a different sample of QSO data than does \cite{Jain_Dev_Alcaniz_2003}, and my BAO measurements have all been updated relative to those of \cite{Dolgov_Halenka_Tkachev_2014, Shafer_2016, Tutusaus_et_al_2016, Haridasu_AAP_2017}. GRB data were used to constrain the power law ansatz in \cite{Haridasu_AAP_2017}, and many, but not all, of these data are the same as those I use here (additionally, my data set is larger and contains newer measurements). To my knowledge, HIIG data have never been used to constrain power law expansion. Because these data are independent of the $H(z)$, BAO, QSO, and GRB data sets, I obtain tight constraints on the parameters of the power law ansatz when I fit it to these data in combination with the $H(z)$, BAO, QSO, and GRB data (see Sec. \ref{sec:Results}).

\section{Methods}
\label{sec:Methods}

The methods that I use to compare the expansion histories predicted by the power law ansatz and the \lcdm\ model are largely the same as the methods used in \cite{Cao_Ryan_Ratra, Cao_Ryan_Ratra_2021, Cao_Ryan_Khadka_Ratra, Rani_et_al_2015, Ryan_Evidence, Ryan_2, Ryan_1, Shafer_2016, Tutusaus_et_al_2016}, which I briefly summarize here. For each combination of data that I study, I compute the quantity
\begin{equation}
    \chi^2_{\rm min} := -2{\rm ln}\mathcal{L}_{\rm max}
\end{equation}
where the likelihood function $\mathcal{L}$ depends on the parameters of the expansion history under consideration. The likelihood function takes a different form depending on the data combination that is used to compute it; these forms are described in \cite{Cao_Ryan_Khadka_Ratra} and \cite{Cao_Ryan_Ratra}.\footnote{Some of the BAO data that I use are correlated, so it is necessary to take their covariance matrices into account when computing $\chi^2_{\rm min}$. See \cite{Cao_Ryan_Ratra_2021} and \cite{Ryan_2} for the covariance matrices of the correlated data.} For expansion histories having the same number of parameters, the best-fitting history is that which has a smaller value of $\chi^2_{\rm min}$. As in \cite{Cao_Ryan_Ratra, Cao_Ryan_Khadka_Ratra}, I use the \textsc{Python} module \textsc{emcee} \cite{Foreman-Mackey_Higg_Lang_Goodman_2013} to sample the likelihood function $\mathcal{L}$, and I use the \textsc{Python} module \textsc{getdist} \cite{Lewis_2019} both to generate the one- and two-dimensional likelihood contours shown in left and right panels of Fig. \ref{fig:ZBQGH_marginalized} and to compute the one-dimensional marginalized best-fitting values (sample means) and 68\% uncertainties (two-sided limits) of the model/ansatz parameters.

When comparing expansion histories with different numbers of parameters, the $\chi^2$ function is not necessarily the most informative statistic to use, because it gives simple and complex expansion histories equal weight. For this reason, I also use the corrected Akaike Information Criterion:
\begin{equation}
    {\rm AICc} := {\rm AIC} + \frac{2n(n+1)}{N - n - 1},
\end{equation}
where
\begin{equation}
    {\rm AIC} := \chi^2_{\rm min} + 2n,
\end{equation}
is the Akaike Information Criterion (suitable in the limit that $N >> n$), and the Bayes Information Criterion:
\begin{equation}
    {\rm BIC} := \chi^2_{\rm min} + n {\rm ln} N,
\end{equation}
\cite{Liddle_2007}. In the equations above, $n$ is the number of parameters and $N$ is the number of data points.\footnote{In previous work \cite{Cao_Ryan_Ratra, Cao_Ryan_Ratra_2021, Cao_Ryan_Khadka_Ratra, Ryan_2, Ryan_1}, my collaborators and I used the AIC and BIC to compare the quality of cosmological model fits to data. Here I use the AICc in place of the AIC because the AICc is more appropriate for smaller data sets (like the $H(z)$ and BAO sets), because it approaches the AIC in the limit that $N$ is large, and to facilitate the comparison of my results with the results of \cite{Shafer_2016, Tutusaus_et_al_2016}, both of which used the AICc in their analyses.} The AICc and BIC punish expansion histories that have a greater number of parameters, favoring those with fewer parameters. In this sense, the AICc and BIC provide a quantitative basis for choosing which expansion history provides the most parsimonious fit to a given set of data.

\section{Results and Discussion}
\label{sec:Results}

\begin{table*}
    \caption{Best-fitting parameters of the power law ansatz.}
    \label{tab:PL_BFP_Om}
    \centering
    \resizebox{\columnwidth}{!}{%
    \begin{tabular}{cccccccccccc}
    \hline
    \hline
    Data type & $H_0$ (km s$^{-1}$ Mpc$^{-1}$) & $\beta$ & $\om$ & $\obh$ & $a$ & $b$ & $\sigma_{\rm ext}$ & $\nu$ & $\chi^2_{\rm min}/\nu$ & AICc & BIC\\
    \hline
         $H(z)$ & 61.92 & 0.9842 & - & - & - & - & - & 29 & 0.5721 & 21.02 & 23.46 \\
         BAO & 89.78 & 0.9206 & 0.6192 & 0.03819 & - & - & - & 7 & 1.513 & 25.26 & 20.18\\
         QSO & 61.83 & 0.9673 & - & - & - & - & - & 118 & 2.991 & 357.1 & 362.6\\
         BAO+QSO & 60.57 & 0.9213 & 0.2030 & 0.07690 & - & - & - & 127 & 2.864 & 372.0 & 383.2\\
         GRB & 72.34 & 0.7530 & - & - & 49.99 & 1.115 & 0.4010 & 114 & 1.138 & 140.3 & 153.7\\
         HIIG & 70.99 & 1.251 & - & - & - & - & - & 151 & 2.725 & 415.5 & 421.5\\
         GRB+HIIG & 70.31 & 1.158 & - & - & 50.12 & 1.157 & 0.4066 & 267 & 2.039 & 554.7 & 572.5\\      
         All Data & 63.06 & 0.9470 & 0.2234 & 0.06706 & 50.16 & 1.144 & 0.4025 & 427 & 2.229 & 966.2 & 994.4\\
    \hline
    \hline
    \end{tabular}%
    }
\end{table*}

The best-fitting values of the parameters of the power law ansatz (namely, those that minimize the $\chi^2$ function), are recorded in columns 2-8 of Table \ref{tab:PL_BFP_Om}. The number of degrees of freedom, 
\begin{equation}
    \nu := N - n
\end{equation}
is recorded in column 9 of this table. Columns 10-12 record, respectively, the minimum value of the reduced $\chi^2$ function, and the minimum values of the AICc and BIC. Similarly, the best-fitting values of the parameters of the \lcdm\ model are recorded in columns 2-7 of Table \ref{tab:LCDM_BFP_Om}, with the number of degrees of freedom in column 8, the minimum value of the reduced $\chi^2$ function in column 9, and the minimum values of the AICc and BIC in columns 10 and 11, respectively.

In Table \ref{tab:1d_BFP_PL}, in columns 2 and 3, I record the sample means and two-sided uncertainties of the marginalized parameters of the power law ansatz (I exclude the parameters $\Omega_{m0}$, $\Omega_{b0}h^2$, $a$, $b$, $\sigma_{\rm ext}$ from this table because they're nuisance parameters for the power law ansatz). In column 4 I record the sample mean and two-sided uncertainties (computed from the sample mean and two-sided uncertainties of $\beta$) of the current value of the deceleration parameter
\begin{equation}
    \label{eq:q}
    q_0 = \frac{1}{\beta} - 1.
\end{equation}
In column 5 I record $\Delta \chi^2_{\rm min}$, which I define as the difference between the value of $\chi^2_{\rm min}$ as computed within the power law ansatz for a given data combination, and the value of $\chi^2_{\rm min}$ as computed within the \lcdm\ model for the same data combination. The relative probabilities $e^{-\rm \Delta AICc/2}$ and $e^{-\rm \Delta BIC/2}$ of the power law ansatz I record in columns 6 and 7, where $\Delta$AICc and $\Delta$BIC are defined in the same way as $\Delta \chi^2_{\rm min}$. In columns 2 and 3 of Table \ref{tab:1d_BFP_LCDM}, I record the sample means and two-sided uncertainties of the marginalized parameters of the \lcdm\ model, excluding the nuisance parameters $\Omega_{b0}h^2$, $a$, $b$, and $\sigma_{\rm ext}$. In column 4 of Table \ref{tab:1d_BFP_LCDM} I record the sample mean and two-sided uncertainties (computed from the sample mean and two-sided uncertainties of $\Omega_{m0}$) of the current value of the deceleration parameter
\begin{equation}
    q_0 = \frac{\Omega_{m0}}{2} - \Omega_{\Lambda} = \frac{3}{2}\Omega_{m0} - 1.
\end{equation}
The prior probabilities of all parameters are flat, and non-zero within the ranges $20$ km s$^{-1}$ Mpc$^{-1}$ $\leq H_0 \leq 100$ km s$^{-1}$ Mpc$^{-1}$, $0.25 \leq \beta \leq 4$, $0.1 \leq \Omega_{m0} \leq 0.7$, $0.005 \leq \Omega_{b0}h^2 \leq 0.1$, $40 \leq a \leq 60$, $0 \leq b \leq 5$, and $0 \leq \sigma_{\rm ext} \leq 10$.

\begin{table*}
    \caption{Best-fitting parameters of the \lcdm\ model.}
    \label{tab:LCDM_BFP_Om}
    \centering
    \resizebox{\columnwidth}{!}{%
    \begin{tabular}{ccccccccccc}
    \hline
    \hline
    Data type & $H_0$ (km s$^{-1}$ Mpc$^{-1}$) & $\om$ & $\obh$ & $a$ & $b$ & $\sigma_{\rm ext}$ & $\nu$ & $\chi^2_{\rm min}/\nu$ & AICc & BIC\\
    \hline
         $H(z)$ & 68.15 & 0.3196 & - & - & - & - & 29 & 0.5000 & 18.93 & 21.37\\
         BAO & 74.01 & 0.2967 & 0.03133 & - & - & - & 8 & 1.124 & 18.43 & 16.19\\
         QSO & 68.69 & 0.3154 & - & - & - & - & 118 & 2.983 & 356.1 & 361.6\\
         BAO+QSO & 69.51 & 0.2971 & 0.02459 & - & - & - & 128 & 2.821 & 367.3 & 375.7\\
         GRB & 75.65 & 0.7000 & - & 49.98 & 1.108 & 0.4012 & 114 & 1.141 & 140.6 & 154.0\\
         HIIG & 71.81 & 0.2756 & - & - & - & - & 151 & 2.720 & 414.8 & 420.8\\
         GRB+HIIG & 71.45 & 0.2950 & - & 50.17 & 1.136 & 0.4035 & 267 & 2.031 & 552.5 & 570.3\\
         All Data & 70.07 & 0.2949 & 0.02542 & 50.19 & 1.135 & 0.4040 & 428 & 2.148 & 931.6 & 955.8\\
    \hline
    \hline
    \end{tabular}%
    }
\end{table*}

The two-dimensional confidence contours and one-dimensional likelihoods of the power law ansatz, for several combinations of data, are shown in the left panel of Fig. \ref{fig:ZBQGH_marginalized}. The contours and likelihoods associated with the $H(z)$ data are shown as dotted blue curves, those associated with the BAO + QSO data combination are shown as dash-dotted red curves, those associated with the GRB + HIIG combination are shown as dashed green curves, and those associated with the combination of all the data are shown as solid black curves (I combine the standard ruler and standard candle data in these plots to reduce visual clutter). The right panel of Fig. \ref{fig:ZBQGH_marginalized} shows the two-dimensional confidence contours and one-dimensional likelihoods of the \lcdm\ model, for the same data combinations.

From the marginalized parameter fits in Table \ref{tab:1d_BFP_PL}, I find that the best-fitting value of $\beta$ from the $H(z)$, QSO, GRB, and HIIG data is consistent with $\beta = 1$ to within 1-2$\sigma$, in agreement with many of the studies quoted in Table \ref{tab:beta_fits}. This translates to the best-fitting value of $q_0$ being within 1-2$\sigma$ of $q_0 = 0$ for each of these data sets, consistent with a coasting universe. The BAO data, however, are not consistent with $\beta = 1$, the best-fitting value of $\beta$ for this data set being more than 4$\sigma$ away from unity. This means, as reflected in the best-fitting $q_0$ value, that when the power law ansatz is fitted to the BAO data, these data favor a slowly decelerating universe (rather than a coasting one) to more than 4$\sigma$. The BAO + QSO combination also favors a slowly decelerating universe to more than 4$\sigma$. When these data are combined with the $H(z)$, GRB, and HIIG data, the error bars on $\beta$ and $q_0$ tighten, and the central values of these parameters move slightly closer to $\beta = 1$ and $q_0 = 0$, respectively, though the best-fitting value of $q_0$ is still inconsistent with a coasting universe to more than 3$\sigma$ (see also Fig. \ref{fig:ZBQGH_marginalized}).

From Tables \ref{tab:PL_BFP_Om} and \ref{tab:LCDM_BFP_Om}, we can see that the best-fitting power law ansatz has greater $\chi^2/\nu$, AICc, and BIC values than the best-fitting \lcdm\ model across all data combinations, except when these models are fitted to GRB data alone. In this case, power law expansion provides a slightly better fit to the data compared to the expansion history predicted by \lcdm. When we examine the relative probabilities $e^{-\Delta{\rm AICc}/2}$ and $e^{-\Delta{\rm BIC}/2}$ in Table \ref{tab:1d_BFP_PL}, we find that the power law ansatz produces a slightly better fit to the GRB data than does \lcdm. This preference for the power law ansatz over the \lcdm\ model is unique to the GRB data, however, as all other data combinations favor the \lcdm\ model, with the relative probability of power law expansion ranging from a high of $0.7047$ (HIIG data) to a low of $4.151 \times 10^{-9}$ (full data set). 

\begin{table*}
    \caption{Marginalized best-fitting parameters and model comparison statistics for the power law ansatz. The BAO data alone do not place a tight upper limit on the best-fitting value of $H_0$, and the GRB data do not constrain $H_0$ at all, so these limits are omitted from the table.}
    \label{tab:1d_BFP_PL}
    \centering
    \resizebox{\columnwidth}{!}{%
    \begin{tabular}{ccccccc}
    \hline
    \hline
    Data type & $H_0$ (km s$^{-1}$ Mpc$^{-1}$) & $\beta$ & $q_0$ & $\Delta\chi^2_{\rm min}$ & $e^{-\Delta{\rm AICc}/2}$ & $e^{-\Delta{\rm BIC}/2}$\\
    \hline
         $H(z)$ & $62.46^{+2.693}_{-2.694}$ & $1.013^{+0.06983}_{-0.1038}$ & $-0.01283^{+0.1012}_{-0.06805}$ & 2.090 & 0.3517 & 0.3517\\
         BAO & $72.68_{-8.788}$ & $0.9211^{+0.01653}_{-0.01652}$ & $0.08566_{-0.01948}^{+0.01947}$ & 1.595 & 0.03288 & 0.1360\\
         QSO & $63.22^{+4.088}_{-4.091}$ & $1.045^{+0.1142}_{-0.2054}$ & $-0.04306_{-0.1046}^{+0.1881}$ & 0.9498 & 0.6065 & 0.6065\\
         BAO+QSO & $60.60 \pm 1.108$ & $0.9219^{+0.01645}_{-0.01646}$ & $0.08472_{-0.01946}^{+0.01937}$ & 2.626 & 0.09537 & 0.02352\\
         GRB & - & $0.8707_{-0.2782}^{+0.1197}$ & $0.1485^{+0.3670}_{-0.1579}$ & -0.3534 & 1.162 & 1.162\\
         HIIG & $71.30_{-1.814}^{+1.813}$ & $1.310_{-0.1988}^{+0.1219}$ & $-0.2366^{+0.1158}_{-0.07103}$ & 0.6886 & 0.7047 & 0.7047\\
         GRB+HIIG & $70.58 \pm 1.755$ & $1.199^{+0.1016}_{-0.1535}$ & $-0.1660_{-0.07067}^{+0.1068}$ & 2.209 & 0.3329 & 0.3329\\
         All Data & $63.11^{+0.7886}_{-0.7890}$ & $0.9466^{+0.01593}_{-0.01594}$ & $0.05641_{-0.01778}^{+0.01779}$ & 32.53 & $3.067 \times 10^{-8}$ & $4.151 \times 10^{-9}$\\
    \hline
    \hline
    \end{tabular}%
    }
\end{table*}

It is interesting that the best case to be made for power law expansion comes from the standard candle data, as the GRB data favor the power law ansatz and the HIIG data do not strongly disfavor it. In a similar fashion, \cite{Dolgov_Halenka_Tkachev_2014}, \cite{Rani_et_al_2015}, and \cite{Sethi_Dev_Jain_2005} find that standard candle data (in the form of SN Ia measurements) alone do not rule out or strongly disfavor power law expansion. However, when the GRB and HIIG data are combined, with each other and with the cosmic chronometer and standard ruler data, it is the \lcdm\ model that comes out on top. Cosmic chronometer ($H(z)$) data alone also do not favor power law expansion, and neither does the standard ruler (BAO + QSO) combination. Of these three data sets, the QSO set has the least discriminating power, perhaps because of the wide dispersion of the measurements it contains (see the lower left panel of Fig. \ref{fig:Hz/1+z}); as with the HIIG data, power law expansion is not strongly ruled out by QSO data alone. The BAO data have the most discriminating power of any solo data set, the fit of the power law ansatz to these data having the smallest relative probabilities compared to \lcdm. This is also true of the standard candle set (BAO + QSO), which gives a smaller relative probability than either the cosmic chronometer or standard candle (GRB + HIIG) set when the power law ansatz is fitted to this data combination. When the power law ansatz is fitted to the full data set, the relative probabilities decrease drastically, to the point that power law expansion appears to be very strongly ruled out, at $z \lesssim 8$, in favor of the \lcdm\ model. These results are in broad agreement with the findings of \cite{Rani_et_al_2015, Shafer_2016, Tutusaus_et_al_2016, Haridasu_AAP_2017}, although they differ somewhat in the details. In particular, using a set of $H(z)$ data that is slightly different from mine, \cite{Rani_et_al_2015} finds much stronger evidence against power law expansion than I do.\footnote{They quote $\chi^2_{\rm min}/\nu = 1.8131$ for the fit of the power law ansatz to their $H(z)$ data, and $\chi^2_{\rm min}/\nu = 0.7174$ for the fit of the \lcdm\ model to these data, for a difference of $\Delta \chi^2_{\rm min}/\nu = 1.096$. For these same models, I find only $\Delta \chi^2_{\rm min}/\nu = 0.0721$} Both \cite{Shafer_2016} and \cite{Tutusaus_et_al_2016} use BAO data (a smaller set than mine) to test power law expansion. Contrary to my results, the BAO measurements they use favor power law expansion, although they both find that it is strongly disfavored when BAO data are combined with independent probes (SN Ia in \cite{Shafer_2016, Tutusaus_et_al_2016} and SN Ia + CMB in \cite{Tutusaus_et_al_2016}). Using $H(z)$ + BAO + SNe Ia + GRB data, \cite{Haridasu_AAP_2017} also find that power law expansion is strongly ruled out in favor of \lcdm\ ($\Delta$BIC = 28.02), though their combined data set prefers a slightly larger value of $\beta$ ($1.08 \pm 0.04$) than my combined data set, with larger error bars.

\begin{table*}
    \centering
    \caption{Marginalized best-fitting parameters of the \lcdm\ model. The BAO data alone do not place a tight upper limit on the value of $H_0$, and the GRB data do not constrain $H_0$ at all, so these limits are excluded from the table.}
    \begin{tabular}{cccc}
    Data type & $H_0$ & $\Omega_{m0}$ & $q_0$\\
    \hline
    \hline
         $H(z)$ & $67.73^{+3.078}_{-3.077}$ & $0.3323^{+0.04983}_{-0.06988}$ & $-0.5016^{+0.07474}_{-0.1048}$\\
         BAO & $83.47_{-4.272}$ & $0.2982^{+0.01547}_{-0.01771}$ & $-0.5527^{+0.2321}_{-0.02657}$\\
         QSO & $67.28^{+4.901}_{-5.039}$ & $0.3642^{+0.08152}_{-0.1503}$ & $-0.4537^{+0.1223}_{-0.2254}$\\
         BAO+QSO & $69.58 \pm 1.379$ & $0.2975^{+0.01529}_{-0.01746}$ & $-0.5538^{+0.02294}_{-0.02619}$\\
         GRB & - & $0.4767_{-0.07217}$ & $-0.2850_{-0.10826}$ \\
         HIIG & $71.70^{+1.819}_{-1.820}$ & $0.2893^{+0.05099}_{-0.07016}$ & $-0.5661^{+0.07649}_{-0.1052}$ \\
         GRB+HIIG & $71.41^{+1.794}_{-1.795}$ & $0.3073^{+0.05140}_{-0.07055}$ & $-0.5391^{+0.07710}_{-0.1058}$ \\
         All Data & $70.13 \pm 0.9590$ & $0.2943^{+0.01368}_{-0.01523}$ & $-0.5586^{+0.2052}_{-0.2284}$ \\
    \hline
    \hline
    \end{tabular}
    \label{tab:1d_BFP_LCDM}
\end{table*}

That power law expansion is ruled out in favor of the \lcdm\ model, from an analysis of $H(z)$, BAO, QSO, GRB, and HIIG data, is a strong statement, and should not be accepted uncritically. Though I believe I have made a good case against the power law ansatz, a few caveats must also be mentioned:

\begin{figure*}
\caption{The left panel shows one- and two-dimensional constraints on the parameters of the power law ansatz from several combinations of data, and the right panel shows one- and two-dimensional constraints on the \lcdm\ model from the same combinations of data (nuisance parameters excluded).}
\label{fig:ZBQGH_marginalized}
\centering
    \resizebox{\columnwidth}{!}{%
    \includegraphics[scale=1]{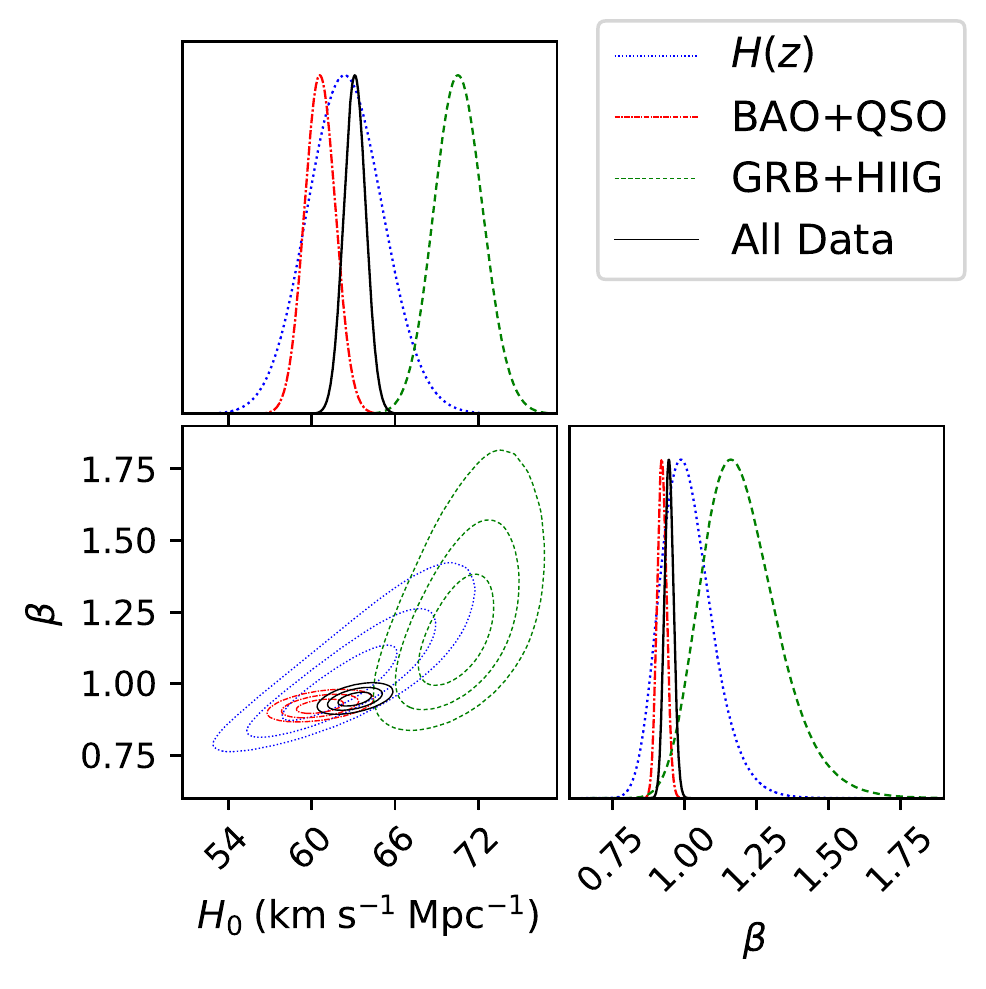}
    \includegraphics[scale=1]{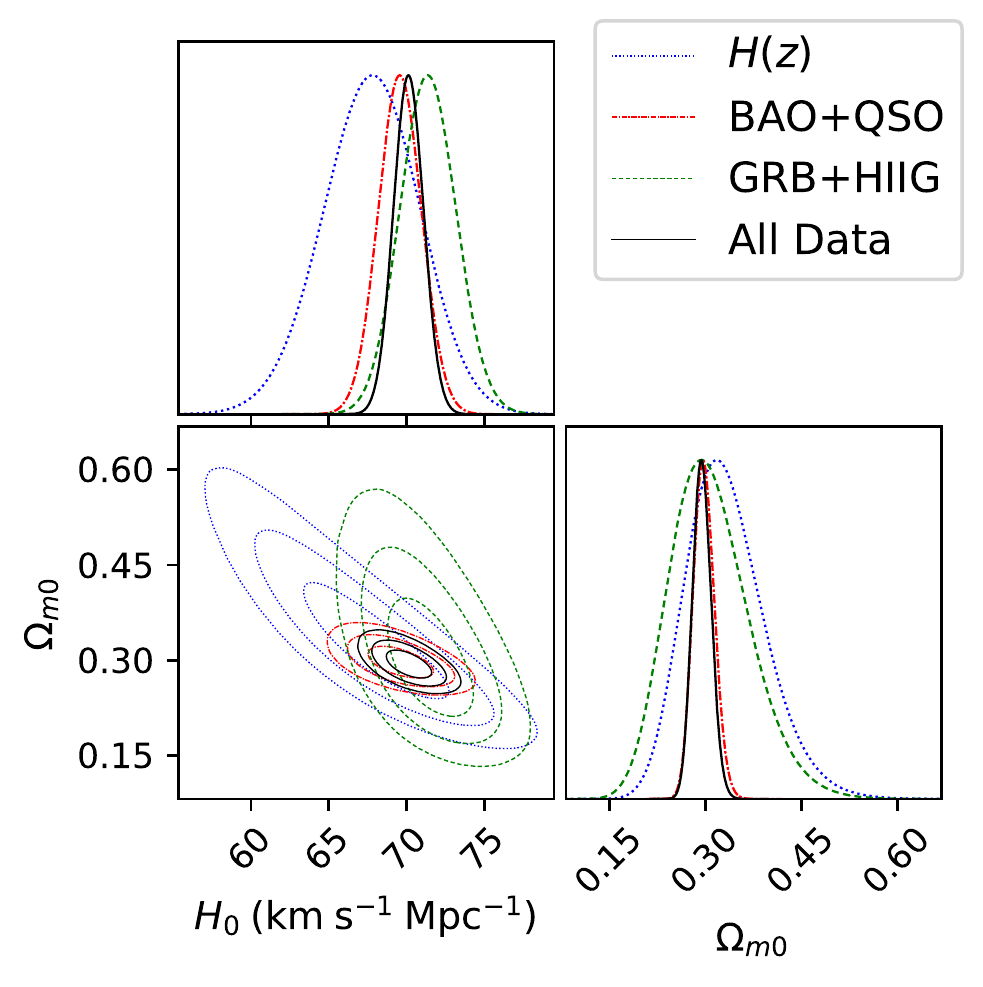}%
    }
\end{figure*}

1.) The results that are shown in Tables \ref{tab:PL_BFP_Om}-\ref{tab:1d_BFP_LCDM} do not take the finite detection significance of the BAO data into account. As discussed in \cite{Ruiz_et_al_2012, Shafer_2016, Tutusaus_et_al_2016}, for a weak BAO signal, one must account for the possibility that the BAO feature in the large-scale matter power spectrum has not actually been detected. To do this, one must replace the standard gaussian $\chi^2$ function $\chi^2_{\rm G} := -2{\rm ln}\mathcal{L}_{\rm G}$ with
\begin{equation}
\label{eq:finite_chi2}
    \chi^2 := \frac{\chi^2_{\rm G}}{\sqrt{1 + \left(\frac{S}{N}\right)^{-4}\chi^4_{\rm G}}},
\end{equation}
where $S/N$, the signal-to-noise ratio, is the detection significance of the BAO feature. As described in \cite{Cao_Ryan_Ratra_2021}, three of the BAO measurements I use in this paper are uncorrelated, and the rest are correlated. To test the robustness of my results, I replaced the gaussian likelihoods of the uncorrelated BAO measurements with their counterparts defined by eq. (\ref{eq:finite_chi2}) and performed the BAO analysis again. I found no significant change in the results when I did this, which is perhaps not surprising; in Fig. 11 of \cite{Ruiz_et_al_2012}, the authors show that accounting for the finite detection significance of BAO data only has the effect of widening the confidence contours (primarily the $3\sigma$ contour) a little, and that this widening almost disappears when BAO data are combined with other probes. With that said, I did not investigate the effect of the detection significance of the correlated BAO data on the model fits, and I do not know how large the effect is for these data. However, based on the above considerations as they apply to the uncorrelated BAO data, I do not expect the effect of the detection significance of the uncorrelated BAO data to be a significant factor affecting the validity of my results.

2.) The fits to the QSO and HIIG data give reduced $\chi^2$ values that are all $>2$. The fit to the $H(z)$ data gives, for both the power law ansatz and \lcdm\ model, reduced $\chi^2$ values comparable to 0.5. The larger reduced $\chi^2$ values suggest that neither the power law ansatz nor the \lcdm\ model is a particularly good fit to the QSO or the HIIG data (though the reduced $\chi^2$ values of the \lcdm\ model are consistently lower for these data than those of the power law ansatz), or that the uncertainties of these data have been underestimated, or both.\footnote{As in \cite{Cao_Ryan_Ratra, Cao_Ryan_Khadka_Ratra, Cao_Ryan_Ratra_2021}, I only consider the statistical errors of the HIIG data. The systematic uncertainties of the HIIG data are the subject of an ongoing investigation by Roberto Terlevich and his colleagues, the results of which will be published in a forthcoming paper (Roberto Terlevich, private communication, 2021).} The reduced $\chi^2$ values of the $H(z)$ data suggest, on the other hand, that the uncertainties of these data have been overestimated. The possible overestimation of the $H(z)$ uncertainties has previously been noted by myself and my collaborators (see \cite{Cao_Ryan_Khadka_Ratra}), and is perhaps apparent in Fig. \ref{fig:Hz/1+z}. My collaborators and I have also previously noted the possible underestimation of the QSO and HIIG error bars \cite{Cao_Ryan_Ratra, Ryan_2}. That the power law ansatz has consistently higher values of $\chi^2_{\rm min}$ for all data sets (GRB excepted) alone and in combination (with the measurements presumably having mostly independent systematics), argues against its validity as a description of cosmic expansion for $z \lesssim 8$, though the argument could be made stronger with a better understanding of the error bars on the measurements.

3.) The $H(z)$ data are somewhat correlated with the QSO data. These data are correlated because some cosmic chronometer data were used to obtain the characteristic angular size $l_m$ of the QSO data. As described in \cite{Cao_et_al2017b}, using the Gaussian Process method \cite{Seikel_Clarkson_Smith_2012}, 24 $H(z)$ measurements at $z \leq 1.2$ were interpolated to produce a cosmological model independent Hubble parameter function $H(z)$. This function was then integrated to produce the angular diameter distances used, in conjunction with angular size measurements $\theta_{\rm obs}$, to obtain $l_m = 11.03 \pm 0.25$ pc. This correlation has been noted in the literature \cite{Cao_Ryan_Ratra_2021} and I currently believe that the parameter constraints from QSO data alone are wide enough that the correlation between these data and $H(z)$ data is not significant. With that said, the magnitude of this correlation is not currently known in detail, and a defender of power law expansion could point to this as a weakness of my study. One could solve this problem by treating $l_m$ as a free parameter in the cosmological model fits, although this tends to produce parameter constraints that are so wide as to be nearly uninformative.\footnote{Shulei Cao, private communication, 2021.}

\section{Conclusion}
\label{sec:Conclusion}

In this paper, I analyzed a set of cosmic chronometer ($H(z)$), standard ruler (BAO and QSO), and standard candle (GRB and HIIG) data to find out whether simple power law expansion (with a constant exponent $\beta$) fits these data as well as or better than the expansion history predicted by the standard \lcdm\ model. Using simple model comparison statistics, similar to what I and many others have used to test alternatives to the \lcdm\ model, I found that the power law ansatz does not provide a good fit to the data when compared to \lcdm. Any cosmological model that predicts power law expansion with a constant exponent is therefore not a viable candidate to replace the \lcdm\ model at $z \lesssim 8$.

My results are consistent with, and complementary to, other recent studies which have investigated how well power law expansion fits low redshift data. These results, along with the constraints set by primordial nucleosynthesis, show that the simple power law ansatz does not adequately describe the evolution of the Universe over the course of its history. 

\section{Acknowledgments}
I thank Shulei Cao, Narayan Khadka, Lado Samushia, and Bharat Ratra for their helpful comments on an early draft of this paper. This work was partially funded by Department of Energy grant DE-SC0011840. The computing for this project was performed on the Beocat Research Cluster at Kansas State University, which is funded in part by NSF grants CNS-1006860, EPS-1006860, EPS-0919443, ACI-1440548, CHE-1726332, and NIH P20GM113109.

\appendix
\section{Direct comparison of expansion histories to data}

\begin{figure*}
    \caption{In all panels, the abbreviation ``PL'' denotes the power law ansatz and $h := H_0/(100\hspace{1mm}{\rm km}\hspace{1mm}{\rm s}^{-1}\hspace{1mm}{\rm Mpc}^{-1})$. In the lower right panel the HIIG data are represented by blue dots, and each GRB datum is represented by a purple ``x''.}
    \label{fig:Hz/1+z}
    \centering
    \resizebox{\columnwidth}{!}{%
    \includegraphics[scale=1]{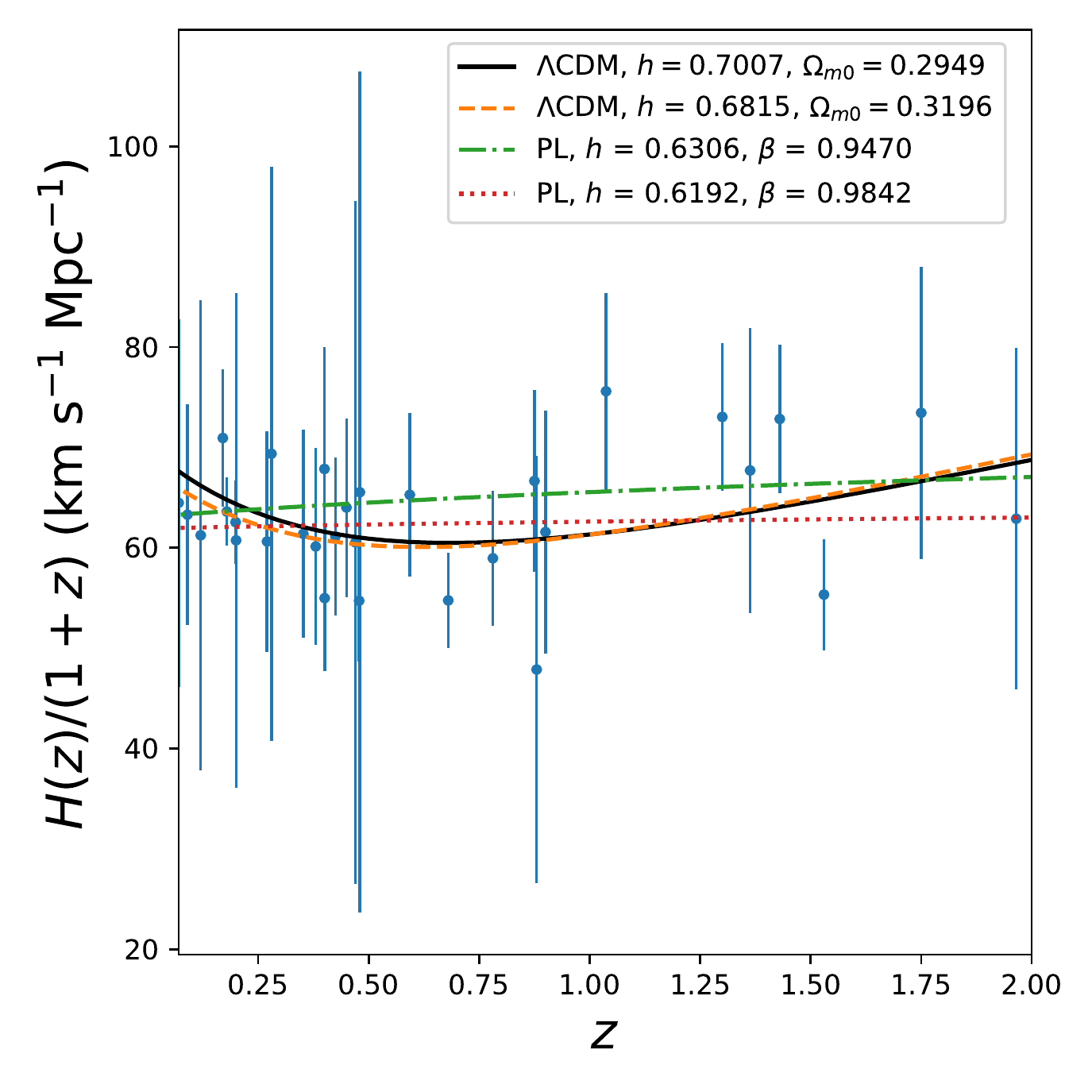}
    \includegraphics[scale=1]{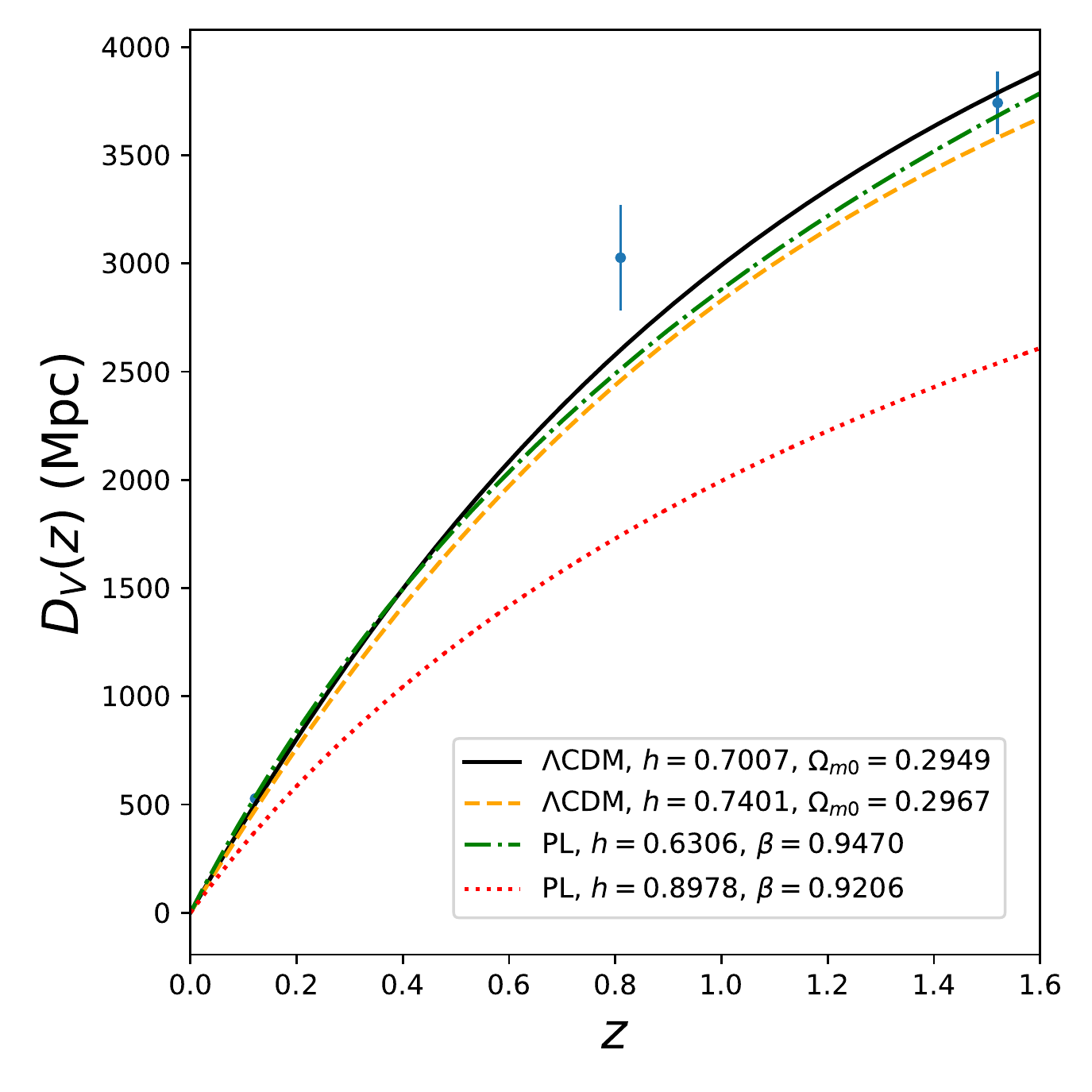}
    }
    \centering
    \resizebox{\columnwidth}{!}{%
    \includegraphics[scale=1]{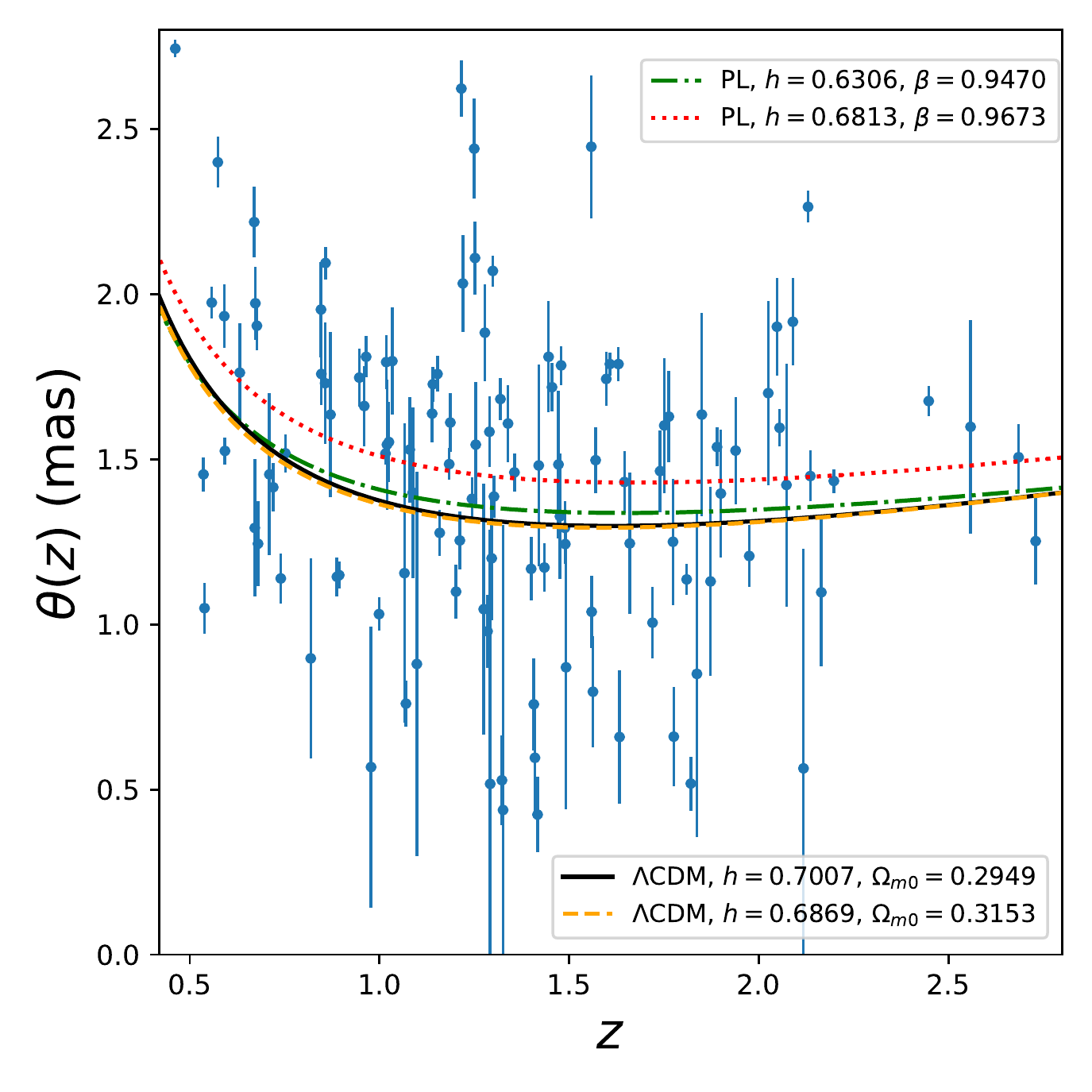}
    \includegraphics[scale=1]{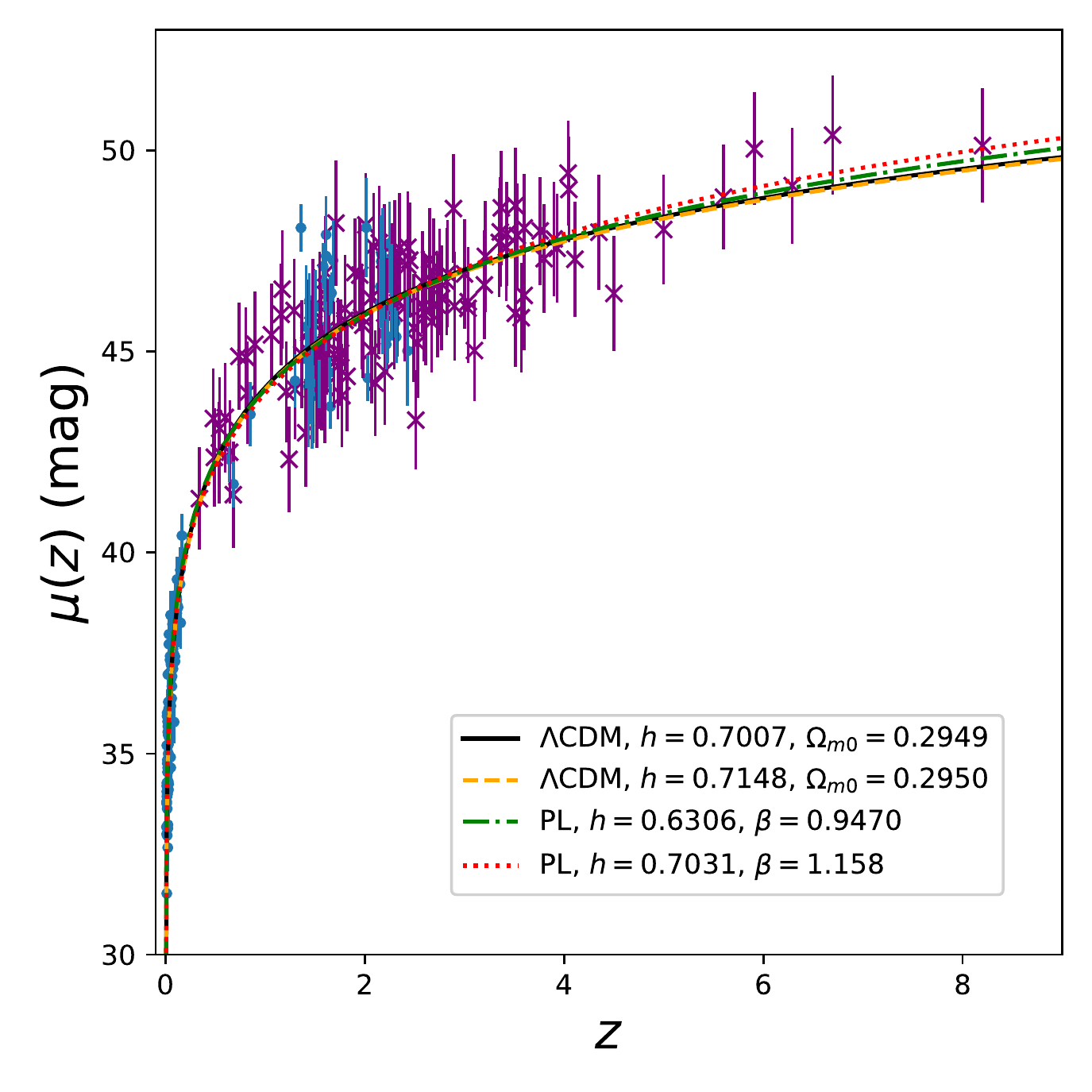}%
    }
\end{figure*}

Here I plot the predicted expansion histories of the power law ansatz and the \lcdm\ model together with the various data sets I use. In the upper left panel of Fig. \ref{fig:Hz/1+z} I plot $\frac{H(z)}{1 + z}$ versus $z$, where the blue dots represent the $H(z)$ measurements and the curves represent the predicted value of $\frac{H(z)}{1 + z}$, as a function of redshift, for the power law ansatz and the \lcdm\ model when these are fitted either to the full data set or to the $H(z)$ data alone. From the figure, we can see that the power law ansatz fails to account for deceleration-acceleration transition which occurs around $z \sim 0.75$.\footnote{For a discussion of the deceleration-acceleration transition, see e.g. \cite{Farooq_Ranjeet_Crandall_Ratra_2017}.} This is backed up by the analyses of \cite{Kumar_2012} and \cite{Rani_et_al_2015} (although a stronger case for this could be made using $H(z)$ data with smaller error bars). 

The upper right panel of Fig. \ref{fig:Hz/1+z} shows a plot of the measured value of the volume-averaged angular diameter distance $D_{\rm V}(z)$, at three different redshifts, from the uncorrelated BAO measurements shown in Table 1 of \cite{Cao_Ryan_Ratra_2021}.\footnote{I did not use the correlated measurements because these do not have independent error bars.} To obtain the central value and error bars of $D_{\rm V}(z)$ at $z = 0.81$, I computed
\begin{equation}
    D_{\rm V}(z) = \left[cz(1 + z)^2 \frac{D_{\rm A}(z)^2}{H(z)}\right]^{1/3}
\end{equation}
from the central value and uncertainty of the $D_{\rm A}(z)$ measurement at $z = 0.81$, along with the median central value and median uncertainty of the two $H(z)$ measurements at $z = 0.70$ and $z = 0.90$ from Table 2 of \cite{Ryan_1}.\footnote{Although the sound horizon $r_s$ (which sets the scale of the BAO measurements) is a function of the model or ansatz parameters, I found that both the \lcdm\ model and the power law ansatz predict $r_s = 144.23$ Mpc when the best-fitting parameters of each (from the full data set) are used to compute it. This means that the $D_{\rm V}(z)$ data are effectively model/ansatz-independent.}  The curves shown in the upper right panel of Fig. \ref{fig:Hz/1+z} represent the predicted values of $D_{\rm V}(z)$ for the \lcdm\ model and the power law ansatz. Although power law expansion appears to be ruled out when fitted to the BAO data alone (as it severely under-predicts the values of $D_{\rm V}(z)$ at all redshifts), when it is fitted to the full data set its predictions are nearly indistinguishable (within the error bars of the measurements) from those of the \lcdm\ model. A stronger (or perhaps weaker) case against power law expansion, however, could presumably be made with more independent measurements, as the dispersion of the values of $D_{\rm V}(z)$ can not be readily inferred from such a small data set.

Large dispersion is a particular problem for the QSO data, as the angular size measurements $\theta(z)$ do not show a clear trend with increasing redshift. This, coupled with the fact that the \lcdm\ and power law predictions of the angular size are very similar over the range of the QSO data, means that these data do not clearly favor one over the other (see the lower left panel of Fig. \ref{fig:Hz/1+z}). Similarly, the \lcdm\ and power law predictions of the distance modulus $\mu(z)$ are almost identical over the redshift range containing the HIIG and GRB data. Although these data show a clear trend with increasing redshift, the predictions of the \lcdm\ model and the power law ansatz only begin to diverge around $z \approx 4$, a redshift beyond which most of the data lie. These data therefore, like the QSO data, do not strongly favor one expansion history over the other (see the lower right panel of Fig. \ref{fig:Hz/1+z}).\footnote{To plot the GRB data model- and ansatz-independently, I used values of $a$ and $b$ computed from the average of the (nearly model/ansatz-independent) best-fitting values of those parameters that are listed in the bottom rows of Tables \ref{tab:PL_BFP_Om} and \ref{tab:LCDM_BFP_Om}.}

\bibliographystyle{JHEP}
\bibliography{main_2}

\end{document}